\documentclass{article}
\usepackage[utf8]{inputenc}
\usepackage{url}
\usepackage{graphicx}
\usepackage{multirow}
\usepackage{xspace} 
\usepackage{framed} 
\usepackage{mdframed}
\usepackage{multirow}
\usepackage{makecell}
\usepackage{caption}
\usepackage{subfig}
\usepackage{comment}
\usepackage{listings}
\usepackage{xcolor}

\def\BibTeX{{\rm B\kern-.05em{\sc i\kern-.025em b}\kern-.08em
    T\kern-.1667em\lower.7ex\hbox{E}\kern-.125emX}}

\definecolor{codegreen}{rgb}{0,0.6,0}
\definecolor{codegray}{rgb}{0.5,0.5,0.5}
\definecolor{codepurple}{rgb}{0.58,0,0.82}
\definecolor{backcolour}{rgb}{0.95,0.95,0.92}

\lstdefinestyle{mystyle}{
    backgroundcolor=\color{white},   
    commentstyle=\color{codegreen},
    keywordstyle=\color{magenta},
    numberstyle=\tiny\color{codegray},
    stringstyle=\color{codepurple},
    basicstyle=\ttfamily\footnotesize,
    breakatwhitespace=false,         
    breaklines=true,                 
    captionpos=b,                    
    keepspaces=true,                 
    numbers=left,                    
    numbersep=2pt,                  
    showspaces=false,                
    showstringspaces=false,
    showtabs=false,                  
    tabsize=2
}

\newcommand{\AHE}[1]{{\color{black}#1}}
\newcommand{\fmc}[1]{{\color{magenta}#1}}

\newcommand{\NT}[1]{{\color{red}#1}}
\newcommand{\daphnesched}[0]{\emph{DaphneSched}}

\title{\daphnesched{}: A Scheduler for \\Integrated Data Analysis Pipelines}

\author{Ahmed Eleliemy and Florina M. Ciorba\\ University of Basel, Switzerland }
\begin{document}
\date{}
\maketitle
\pagenumbering{gobble}

\begin{abstract} 
DAPHNE is a new open-source software infrastructure designed to address the increasing demands of integrated data analysis~(IDA) pipelines, comprising data management~(DM), high performance computing~(HPC), and machine learning~(ML) systems.
Efficiently executing IDA pipelines is challenging due to their diverse computing characteristics and demands. 
Therefore, IDA pipelines executed with the DAPHNE infrastructure require an efficient and versatile scheduler to support these demands.
This work introduces \daphnesched{}, the task-based scheduler at the core of DAPHNE~\cite{damme2022cidr}.
\daphnesched{} is versatile by incorporating eleven task partitioning and three task assignment techniques, bringing the state-of-the-art closer to the state-of-the-practice task scheduling.
To showcase \mbox{\daphnesched{}'s} effectiveness in scheduling IDA pipelines, we evaluate its performance on two applications: a product recommendation system and training of a linear regression model.
We conduct performance experiments on multicore platforms with 20 and 56 cores, respectively.
The results show that the versatility of \daphnesched{} enabled combinations of scheduling strategies that outperform commonly used scheduling techniques by up to 13\%. 
This work confirms the benefits of employing \daphnesched{} for the efficient execution of applications with IDA pipelines.


\end{abstract}

\section{Introduction}
\label{sec:intro}

Scientific applications often combine data analysis pipelines that involve data management~(DM) (data query processing), high-performance computing~(HPC) (large- and multi-scale simulations), and machine learning steps~\cite{DEELMAN201517, MLCompute}.
This diversity requires convergence between DM, HPC, and ML systems.
Today, we see full convergence at the hardware infrastructure level, i.e., data and HPC centers share similar computing, networking, and storage infrastructure, while convergence at the software infrastructure is still an open challenge~\cite{ihde2021survey,cheng2018experiences}. 
DM, HPC, and ML pipelines rely on various data representations, programming paradigms, execution frameworks, and runtime libraries. 
Thus, these pipelines are considered for execution independently which leads to missing many performance optimization opportunities, e.g., zero-data copy and efficient usage of heterogeneous computing resources~\cite{damme2022cidr}.
Therefore, a software infrastructure for integrated data analysis~(IDA) pipelines is highly important and required by the scientific community.  

IDA software infrastructures are challenged by 1) \emph{deployment challenge}: integration of different systems, programming paradigms, resource managers, and data representations, and 2) \emph{heterogeneity challenge} increase of specialization at the device level CPUs, GPUs, FPGAs, and computational memory and storage~\cite{damme2022cidr}. These challenges require efficient scheduling that is \emph{versatile} enough to support a wide range of data representations, application characteristics, and target systems. In this work, we introduce \daphnesched{}, a task-based scheduler for IDA pipelines. 

\daphnesched{} differs from existing runtime schedulers~\cite{lill2021proteomics} as follows.
\AHE{DAPHNE~\cite{DAPHNE-WEBSITE}} is an open and extensible software infrastructure for IDA pipelines. 
Unlike existing runtime systems, such as HPX~\cite{HPX}, StarPU~\cite{StarPU}, Charm++~\cite{charm++}, and TensorFlow~\cite{tensorflow2015}, DAPHNE is an incubator, designed for IDA pipelines with DM, HPC, and ML steps rather than either of such steps from a single domain alone. 
Thus, DAPHNE gives opportunities for holistic optimization by considering data representations and efficient execution plans, including the selection of target devices, work partitioning, work assignment, execution ordering, and execution timing.   
\daphnesched follows two design principles: \emph{Extendability} and \emph{Coverage}. \emph{Extendability} refers to the possibility of extending \daphnesched{} with user-defined scheduling schemes. 
This principle is important to ensure the suitability of \daphnesched{} foreseen use case.
\emph{Coverage} refers to considering the majority of all task-based scheduling schemes from the state-of-the-art and state-of-the-practice. This principle is also crucial, as an IDA scheduler must be versatile enough to cover use cases from various domains. 
IDA schedulers must adapt to different applications and target systems, and no single scheduling technique is universally effective. 

Therefore, a scheduler that supports a wide range of techniques has a higher potential to be suitable for more application-system pairs
For instance, \daphnesched{} covers and supports various scheduling strategies, in addition to the default static scheme (STATIC)~\cite{STATIC}.
Specifically, \daphnesched{} supports various dynamic loop scheduling schemes for work partitioning, such as guided self-scheduling (GSS)~\cite{GSS}, trapezoid self-scheduling (TSS)~\cite{TSS}, fixed-size self-scheduling (FSC)~\cite{FSC}, factoring (FAC)~\cite{FAC}, trapezoid factoring self-scheduling (TFSS)~\cite{TSS}, fixed-increase self-scheduling (FISS)~\cite{FISS}, variable-increase self-scheduling (VISS)~\cite{FISS}, performance loop-based self-scheduling (PLS)~\cite{PLS}, and probabilistic self-scheduling (PSS)~\cite{PSS}. 
For work assignment, \daphnesched{} supports two strategies: 1) self-scheduling from a centralized work queue, and 2) work-stealing from multiple queues (per core or per CPU). 
Moreover, \daphnesched{} supports four victim selection strategies: sequential (SEQ), sequential prioritized (SEQPRI),  random (RND), and random prioritized (RNDPRI) work-stealing~\cite{SEQ}. 

\daphnesched{} enables users to combine work partitioning and assignment techniques in novel ways that have not been explored in previous studies, resulting in improved performance in specific scenarios (shown in Sections~\ref{sec:approach} and~\ref{sec:res}). \\
This work makes the following contributions. 

\textbf{C.1} Design and implementation of an extendable and versatile task-based scheduler that shows performance potential for IDA pipelines compared to existing schedulers.\\  
\indent \textbf{C.2} Proposal to combine work stealing with self-scheduling mechanisms to resolve the well-known challenge of how many tasks an thief worker should steal.

We also present a performance study that shows the advantages of \daphnesched{} for two real test cases.

The remainder of this paper is organized as follows. 
In Section~\ref{sec:rel}, we briefly present the background including the terminology that the reader may need to follow this work, and also highlight the most relevant research efforts to this work. 
We introduce the proposed design of \daphnesched{} in Section~\ref{sec:approach}. 
The experimental evaluation of the proposed scheduler is discussed in Section~\ref{sec:res}. 
We conclude this work by highlighting the performance potential and future work to address current limitations of \daphnesched{} in Section~\ref{sec:con}.

\section{Background and Related Work}
\label{sec:rel}

\begin{figure}
    \centering
    \includegraphics[width=0.8\linewidth]{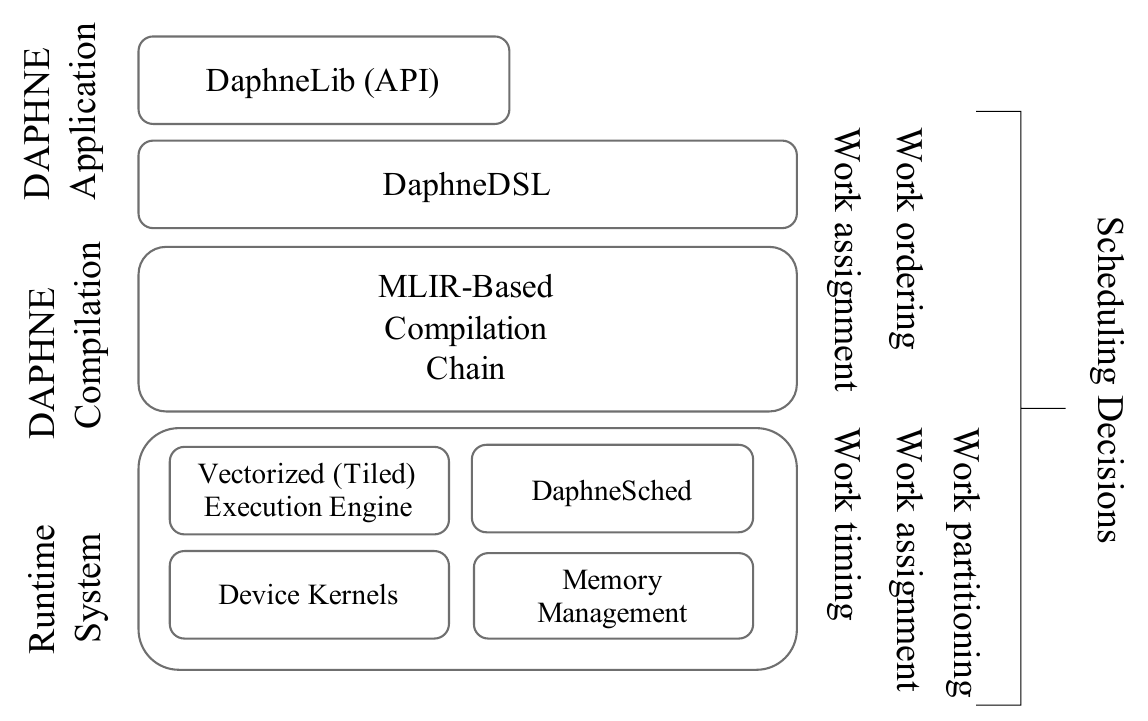}
    \caption{System architecture of the DAPHNE infrastructure. \textbf{Scheduling decisions} are taken by various components at different levels in the DAPHNE system architecture.} 
    \label{fig:systemDesign}
    \vspace{-0.25cm}
\end{figure}

\textbf{Background.} DAPHNE is a scalable infrastructure system designed for integrated data analysis pipelines that encompass data management and processing, as well as high-performance computing and machine learning training and scoring. Figure~\ref{fig:systemDesign} illustrates the DAPHNE's layered system design, comprising four significant components: 
DaphneLib, 
DaphneDSL, 
MLIR-Based Compilation Chain, and 
DaphneRuntime. 
The primary access points to the DAPHNE infrastructure are through DaphneLib and DaphneDSL. 
DaphneLib provides users with interfaces that enable them to make calls to the DAPHNE infrastructure in their Python codes, while 
DaphneDSL is a domain-specific language for ML and numerical computations that is similar to Numpy~\cite{harris2020array}, R~\cite{morandat2012evaluating}, SystemDS~\cite{boehm2019systemds}, and Julia~\cite{bezanson2012julia}. 
The figure shows that scheduling decisions are taken by various components in the DAPHNE architecture.
For instance, the DAPHNE compilation determines the execution order of the operators and type of target devices (CPUs, GPUs, FPGAs, computational storage)~\cite{d51}, while the runtime system determines the mapping (assignment) between work items and specific instances of devices (CPU, GPUs, and FPGAs). 

\textbf{Terminology.} 
The DAPHNE infrastructure is developed by a consortium of various communities, comprising experts in DM, HPC, and ML domains. One of the challenges faced by the consortium was establishing a common understanding of the terminology, as different communities tend to use the same term in different ways. 
We clarify here the definitions and terms used within the context of DAPHNE.
The term \emph{operator} refers to a single, indivisible operation that can be applied to input data. Examples of common operators include matrix operations, such as addition, subtraction, and multiplication.
Operators and data items are the pillars of any computation in the context of DAPHNE, and the combination of the two forms a {task}. 
\emph{Tasks} are the smallest units of work considered for scheduling by the DAPHNE runtime system. 
\emph{Task granularity} is the size of the data item associated with the task.

\daphnesched{} has two independent steps: 1) \emph{work partitioning} that determines tasks' granularity and 2) \emph{work assignment} that determines mapping between workers and tasks. 
Both rely on techniques discussed in the following.

\textbf{Self-scheduling} 
When a worker is free and idle, it will attempt to obtain new work items to execute on its own. The number of work items each worker should \textit{self-schedule} can be determined using various scheduling techniques. 
For example, when the work items are loop iterations without dependencies, dynamic loop self-scheduling (DLS) techniques can be utilized, each technique employing a unique formula to calculate the number of iterations to be self-scheduled each time. 
This amount of loop iterations is referred to as a chunk, and the chunk calculation formulas of the various DLS techniques generate fixed, increasing, decreasing, or random chunk sizes. 

In \daphnesched{}, we employ the following self-scheduling techniques: self-scheduling~(SS)~\cite{SS}, fixed size chunk~(FSC)~\cite{FSC}, guided self-scheduling~(GSS)~\cite{GSS}, trapezoid self-scheduling~(TSS)~\cite{TSS}, factoring~(FAC)~\cite{FAC}, trapezoid factoring self-scheduling~(TFSS)~\cite{TFSS}, fixed increase self-scheduling~(FISS)~\cite{FISS}, variable increase self-scheduling~(VISS)~\cite{FISS}, performance loop-based self-scheduling~(PLS)~\cite{PLS}, and probabilistic self-scheduling(PSS)~\cite{PSS}.
A detailed description of the techniques can be found \AHE{in~\cite{eleliemy2021distributed, penmatsa2007implementation, chronopoulos2007multi}}.


\textbf{Work-stealing and victim selection.} Another strategy for work assignment is work-stealing, which involves a worker attempting to acquire new work items from neighboring workers when it runs out of tasks to perform from its local queue. 
The worker acting as the thief must select a victim worker to steal from. 
Various victim selection strategies are possible. 
In \daphnesched{} we use the following SEQ, SEQPRI, RND, and RNDPRI. 
Sequential victim selection~(SEQ) denotes a worker searching for the victim in a round-robin fashion, starting from their current position in the system topology~\cite{SEQ}.
Sequential prioritized victim selection~(\mbox{SEQPRI}) prioritizes the search for victims within the same NUMA domain. \mbox{SEQPRI} preserves data locality between NUMA domains whenever possible, and minimizes inter-socket communication~\cite{chen2017task}. 
Random victim selection~(RND) involves a random selection of one victim using uniform random distribution over all victims. 
Random prioritized victim selection~(RNDPRI) is similar to RND except that victims are divided according to their NUMA domains, i.e., RNDPRI randomly selects from the victims within the same NUMA domain.


\textbf{Related work.} LB4OMP is a recent load-balancing library that reduces the gap between the literature and practice in the context of DLS techniques~\cite{LB4OMP}. 
LB4OMP extends the LLVM OpenMP runtime library with thirteen DLS techniques. 
LB4OMP and \daphnesched{} overlap in terms of the DLS techniques they both support. 
As LB4OMP is an OpenMP runtime library, it only benefits OpenMP applications, with work sharing loops. 
For task scheduling, the LLVM OpenMP runtime relies on the user to define the granularity of the tasks (i.e., task partitioning) and then uses either RND or RNDPRI as victim selection strategies~\cite{giger2021task}. 
While OpenMP is extensively used in HPC applications, it is less common in DB and ML systems. 
In contrast, \daphnesched{} is part of the DAPHNE runtime system which targets IDA pipelines, comprising HPC, DB, and ML steps. 
The distinguishing feature of \daphnesched{} is allowing a mix of self-scheduling schemes for work partitioning with work-stealing mechanisms for work assignment. 

Fractiling is the first strategy that combines self-scheduling and work-stealing schemes~\cite{Fract}. 
Fractiling determines chunk sizes globally using FAC~\cite{FAC} and chunk shapes locally using tiling, i.e., the computation space is initially divided into tiles to promote locality. 
Faster processors borrow decreasing size sub-tiles of work units from slower processors to balance loads. 
\daphnesched's is similar to Fractiling by separating work partitioning and work assignment. 
However, \daphnesched{} extends Fractiling by supporting additional work partitioning schemes and victim selection strategies.
Specifically, \daphnesched{} allows users to choose any self-scheduling scheme with any of the supported work-stealing mechanisms, i.e., tasks can be generated according to fixed, decreasing, or increasing granularities. Tasks are statically distributed to workers, and once a worker is free and idle, it may steal tasks from neighbors. 
The novelty in \daphnesched{} is that the stolen tasks follow the chosen self-scheduling technique that can be any other supported technique or FAC, similar to Fractiling.

LB4MPI~\cite{LB4MPI1,LB4MPI2} is a research library for scheduling MPI applications. 
LB4MPI and LB4OMP~\cite{LB4OMP} overlap in the set of the DLS techniques they support. 
LB4MPI and LB4OMP~\cite{LB4OMP} provide the same set of DLS techniques, with the main distinction that LB4MPI supports loop scheduling in  MPI applications executing on distributed-memory systems.
To simplify data management, LB4MPI currently assumes the loop data is replicated among workers. 
As LB4MPI is designed for MPI applications, DB and ML applications not using MPI cannot benefit from it.  
Like LB4MPI, \daphnesched{} supports task scheduling on distributed-memory systems. 
As tasks contain both operators and the required data, \daphnesched{} relaxes LB4MPI's current assumption of replicated data.

For HPC applications, there are also multiple task-based runtime technologies, e.g., HPX~\cite{HPX} and StartPU~\cite{StarPU}. 
In these runtime systems, schedulers typically support work-stealing but either use a default static amount of tasks to steal (usually one task) or require the user to set a fixed task granularity to steal throughout the execution. 
In contrast, \daphnesched{} provides a wide range of self-scheduling schemes for work partitioning, allowing tasks of various granularities: increasing, decreasing, and fixed. Thus, whenever a task-stealing mechanism is chosen the granularity of stolen tasks is variable and follows a dynamic partitioning scheme.  

\section{DaphneSched: a Task-based Scheduler for Integrated Data Analysis Pipelines}
\label{sec:approach}

\begin{figure}[!h]
    \centering
    \includegraphics[width=\linewidth]{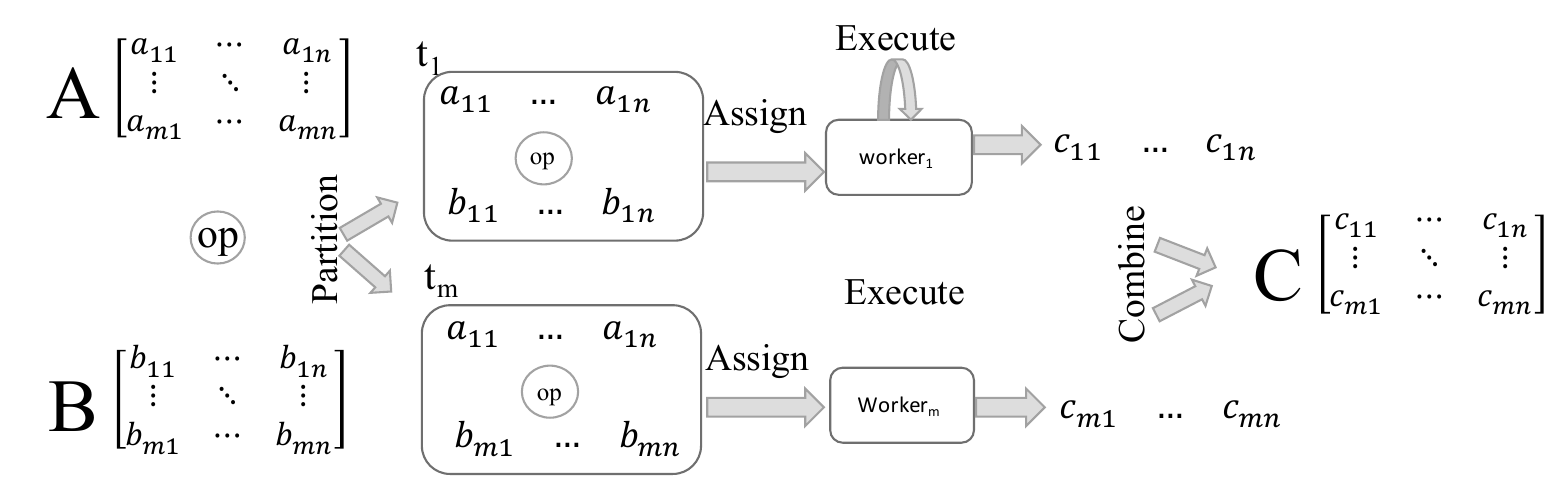}
    \caption{Data parallelism in the DAPHNE runtime system.}
    \label{fig:data-parallelism}
\end{figure}

The DAPHNE infrastructure is designed to support both functional and data parallelism. 
However, the current version of the infrastructure~\cite{DAPHNE-WEBSITE} exploits data parallelism, meaning the  input data is partitioned and the same operation (or small set of operations) is (are) applied to each partition simultaneously, as illustrated in Figure~\ref{fig:data-parallelism}. 

Employing data parallelism to sparse data is challenging as the time to apply one operator depends on 
(1)~the size of the partitioned data, 
(2)~the hardware that executes the operator, and 
(3)~the order of execution that respects spatial data locality. 
\daphnesched{} addresses these challenges as follows.

\begin{figure}
            \centering
    \subfloat[Tasks are of fixed size, with each chunk containing a different number of tasks]
    {\includegraphics[clip,trim=0cm 0cm 0cm 0cm, width=0.40\textwidth, keepaspectratio]{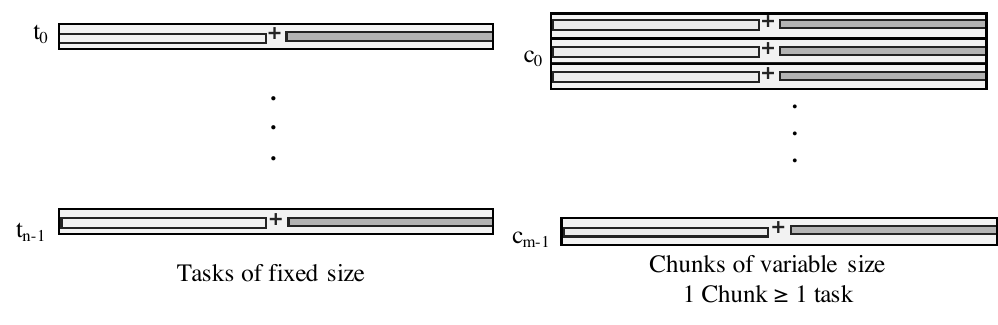}\label{fig:chunks}}
    \hspace{1cm}
    \subfloat[Tasks are of variable sizes, with each chunk containing one task]
    {\includegraphics[clip,trim=0cm 0cm 0cm 0cm, width=0.40\textwidth, keepaspectratio]{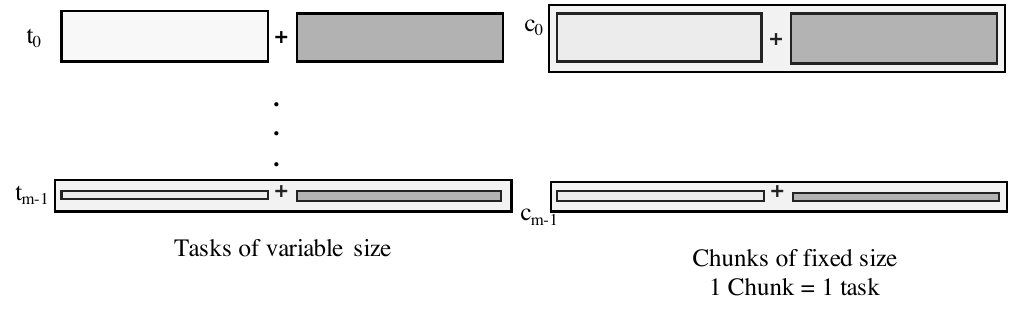}\label{fig:chunks-as-tasks}}
  
        \caption{Work granularity depends on the size of the chunk.}
        \label{fig:all_chunks}   
 \end{figure}

\textbf{From data to tasks.}
\daphnesched{} is a task-based scheduler, that converts inputs from the \AHE{vectorized execution engine~(VEE)~\cite{d41}}, namely data and operators, into tasks. 
As stated in Section~\ref{sec:rel}, tasks are the smallest unit of work to schedule. 
As DAPHNE currently exploits data parallelism, the work granularity or task granularity is dictated by the size of the data. 
Since DAPHNE relies on dense and sparse matrix data structures~\cite{damme2022cidr}, the smallest data size can be one row, one column, or a tile of a certain size. 
In fact, the proper data size should be considered based on the size of the lower levels of the cache of the target system.
For simplicity, one can assume that the smallest data within a task is one row.
The strategy of creating and executing fine-grained tasks minimizes load imbalance, but it comes with a high overhead that increases the execution time. 
One way to address this challenge is to still create fine-grained tasks but to schedule them in chunks of tasks (see Figure~\ref{fig:chunks}). 
This approach reduces the overhead of scheduling individual tasks. 
Another approach is to create tasks of variable sizes (see Figure~\ref{fig:chunks-as-tasks}). 
\daphnesched{} uses the latter approach to avoid the unnecessary level of abstraction (chunks of tasks). 
\daphnesched{} employs the self-scheduling techniques discussed in Section~\ref{sec:rel} to partition the work and determine the task sizes.

\begin{figure}
    \centering
    \includegraphics[width=\textwidth]{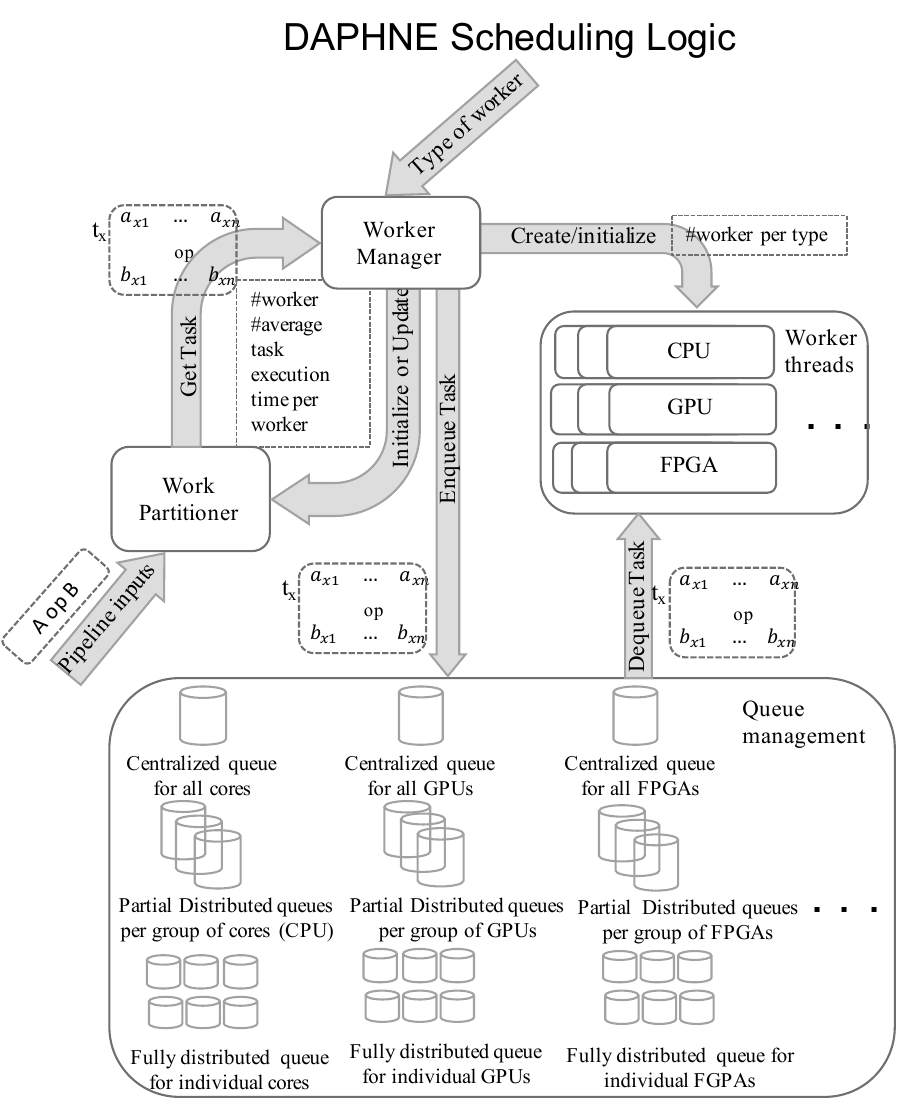}
    \caption{Design and internal working of \daphnesched{}.}
    \label{fig:daphne-sched}
\end{figure}

\textbf{Task partitioning.}
Figure~\ref{fig:daphne-sched} schematically illustrates the work (or task) partitioner that divides tasks into partitions. 
The task partitioner has two interface points: 1)~\texttt{Initialize/Update} which sets the number of workers (threads), the partitioning scheme, and the total number of tasks. 
This point is also used to give updates regarding the runtime information to the partitioner, and 
2)~\texttt{Get Task} which provides a task for execution. The partitioner is intended to be used iteratively to support adaptive scheduling techniques.
Task partitioning can thus be adaptive, based on runtime information. 
\daphnesched{} supports eleven partitioning schemes (STATIC, SS, MFSC, GSS, TFSS, FAC2, TFSS, FISS, VISS, PLS, PSS), discussed in Section~\ref{sec:rel}. 
\AHE{FAC2 and MFSC are two practical implementations of the original FAC~\cite{FAC} and FSC~\cite{FSC}, respectively. Neither of these practical implementations requires prior application profiling data~\cite{LB4OMP}}.

\textbf{Worker management.}
\daphnesched{} is designed to support different types of computing resources, e.g., CPUs, GPUs, FPGAs, and computational storage. 
The DAPHNE compiler decides the type of computing resource that executes a specific pipeline. 
The worker manager initiates workers (threads) that execute or interface with the corresponding devices. \AHE{For instance, in the case of CPU workers, the worker manager creates/manages worker threads that execute the tasks on CPUs.} 
It also creates worker threads that perform data transfers and launch kernels on target devices, such as GPUs and FPGAs.  

\textbf{Queue management.}
The number of generated tasks is often larger than the number of available workers.
\daphnesched{} offers a queuing system that stores tasks until workers become free to execute them.
\daphnesched{} offers multiple queuing strategies, as shown in Figure~\ref{fig:daphne-sched} and described next. 

1)~\textit{Centralized work queue per type of computing resource}: As discussed in Section~\ref{sec:rel}, tasks combine both data and operators to be applied to the data. 
When the DAPHNE compiler decides to map a particular operation to a specific device (CPU, GPU, FPGA), it generates the corresponding code that executes on the chosen device type.
Therefore, currently, \daphnesched{} cannot have a centralized queue per multiple types of computing resources.

2)~\textit{Distributed work queues across individual workers:} Each worker has a separate (local) work queue from which it self-obtains tasks. 
Once its work queue is empty, the worker seeks tasks from other workers. 
The distributed queues across individual workers enable \textit{work-stealing}, \AHE{where a free and idle worker (thief) steals work from other workers (victims)}.

3)~\textit{Distributed work queues across groups of workers:} This type of queue is mainly to support NUMA-domains, i.e., cores that belong to the same NUMA-domain can share the work queue of that domain as the cost of obtaining tasks from this queue is much smaller compared to obtaining tasks from a queue residing on a different NUMA-domain. 

\textbf{Extendability.}
\AHE{The choice of work partitioning scheme has a strong impact on the performance of applications.  For instance, when work stealing is chosen as a work assignment scheme, the granularity of the stolen tasks is determined by the chosen work partitioning scheme.  Thus, the choice of the work partitioning scheme indirectly influences the work assignment schemes.} 
\daphnesched{} is extendable as users are allowed to modify any existing or add any new partitioning scheme. 
To accommodate custom partitioning schemes, DAPHNE developers need to extend the \texttt{getNextChunk} function. 

\begin{figure}
    \centering
    \includegraphics[width=\linewidth]{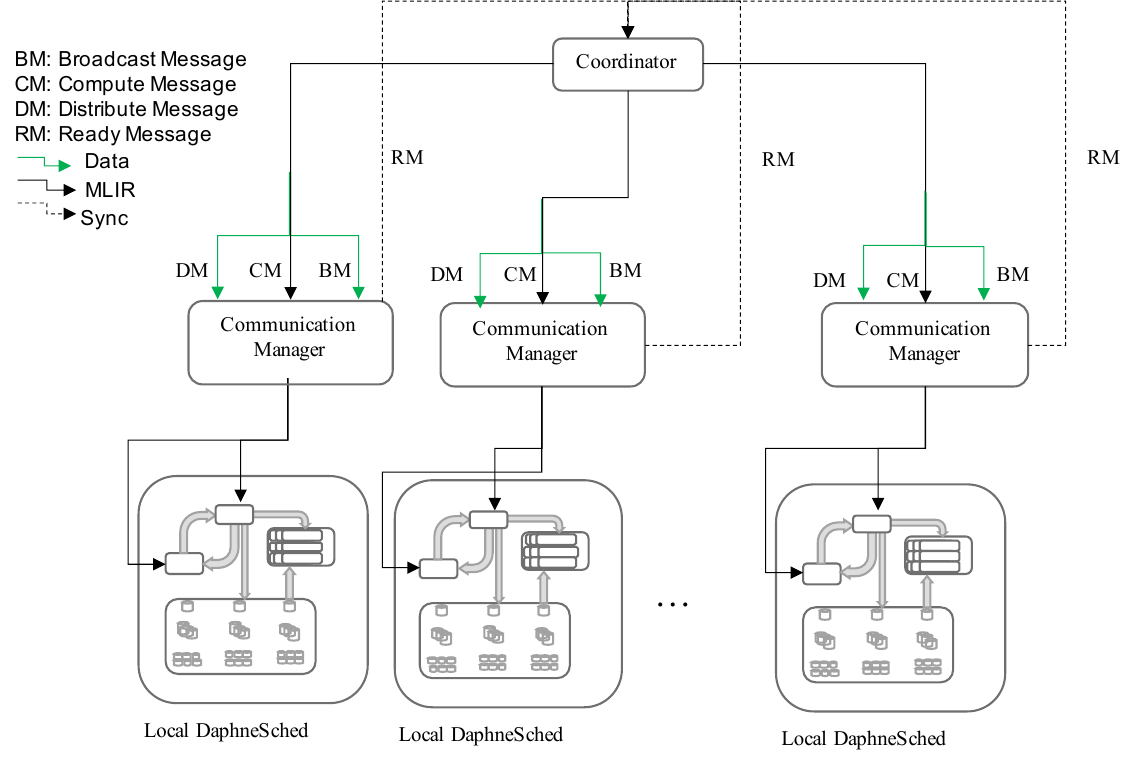}
    \caption{\daphnesched{} for distributed-memory systems.}
    \label{fig:distr}
\end{figure}

\textbf{\daphnesched{} for distributed-memory systems.}
In Figure~\ref{fig:distr}, we see the design extension of DaphneSched to support distributed-memory systems. The design reuses the DaphneSched for shared-memory systems. The distributed DaphneSched adds a new component called coordinator that interfaces with multiple instances of shared-memory DaphneScheds. The coordinator serves as an entry point that the DAPHNE runtime uses, i.e., the DAPHNE runtime system replies to the coordinator to divide, distribute, and collect tasks and results from DaphneSched instances. The DaphneSched experience minor modifications, i.e., they listen to incoming messages from the coordinator. The coordinator sends various messages/data
For instance,  we have 1) distributed pipeline inputs, 2) broadcast pipeline inputs, and 3) MLIR code. On the other side, the worker accepts and stores data items as they come, once the DAPHNE worker gets the MLIR code, it starts to generate local tasks and execute them.

\section{Experimental Evaluation and Discussion}
\label{sec:res}

\begin{figure}
    \centering
    \subfloat[Connected components]
    {\includegraphics[clip,trim=0cm 0cm 0cm 0cm, width=\linewidth, keepaspectratio]{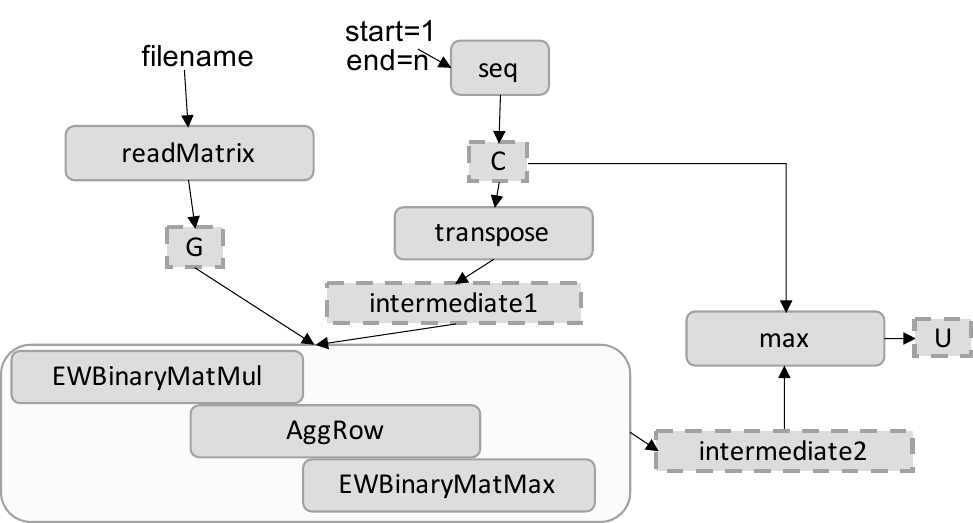}\label{fig:ids-cc}}
  
    \subfloat[Linear regression]
    {\includegraphics[width=\linewidth, keepaspectratio]{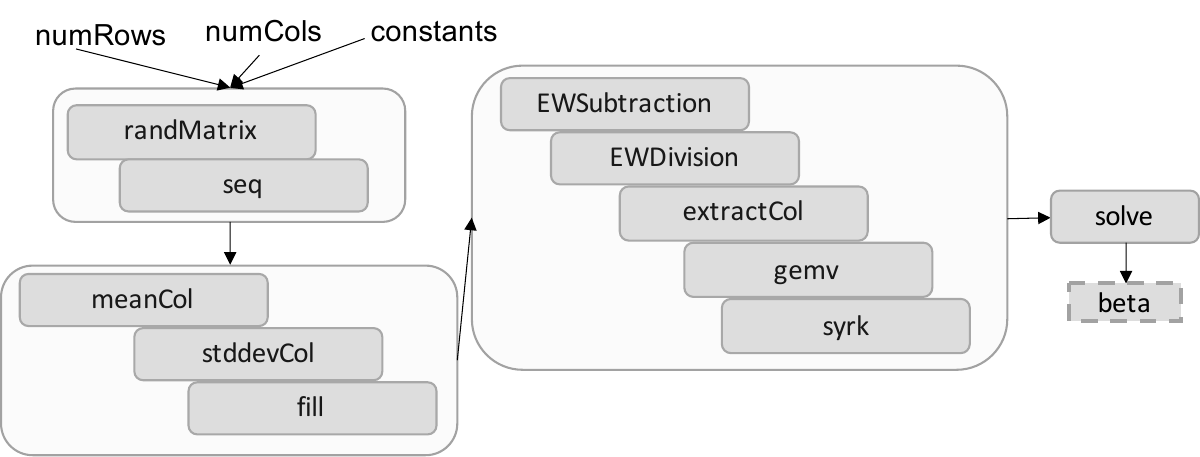}\label{fig:ids-lm}}
    \caption{IDA pipelines as realized by the DAPHNE system.}
    \vspace{-0.3cm}
\end{figure}

\textbf{Applications.} We evaluate \daphnesched{} on two IDA pipelines: 
 
\emph{1) Connected Components.} 
A common product recommendation strategy relies on the frequency of co-purchased items, i.e., whenever product $i$ is frequently purchased together with product $j$. 
The data of the purchased items can be stored as a \emph{sparse matrix} where rows and columns are products and the intersection between the i$^{th}$ row and the j$^{th}$ column is 1 whenever products $i$ and $j$ are co-purchased. 
We use the \emph{Connected Components} algorithm from graph theory to identify co-purchased products. 
The connected components algorithm is implemented in DaphneDSL as shown in Listing~\ref{ls:cc}.
Figure~\ref{fig:ids-cc} shows the different components of the IDA in the connected components algorithm as realized by \daphnesched{}.
\begin{lstlisting}[caption=Connected components algorithm in DaphneDSL~\cite{damme2022cidr}., label={ls:cc}, language=Python]
# Connected components.
# Arguments: - f ... adjacency matrix filename
# Read adjacency matrix.
G = readMatrix($f); 
# Initializations.
n = nrow(G); 
maxi = 100;
c = seq(1, n); 
diff = inf; 
iter = 1;
# Iterative computation.
while(diff>0 & iter<=maxi) {
u = max(rowMaxs(G * t(c)), c); # Neighbor propagation
diff = sum(u != c);  # Changed vertices.
c = u; # Update assignment.
iter = iter + 1;
}
\end{lstlisting}

We use the \emph{Stanford SNAP co-purchasing products of Amazon}\AHE{~\cite{snapnets,leskovec2007dynamics}} as an input data set for the connected components algorithm. 
In this data set, if two products are frequently co-purchased, a directed edge is created between them.
The data set has 403,394 nodes and 3,387,388 edges. A scale-up factor of 50 was applied to the source data set, resulting in an input matrix with 20,169,700 nodes and 244,340,800 two-directional edges~\cite{leskovec2007dynamics}.

\emph{2) Linear Regression} is the most well-known and widely-used type of predictive analysis. 
It is used in various domains to predict the value of a variable based on the value of other known variables. 
Extracting a linear regression model requires solving a linear system of equations. 
The algorithm in Listing~\ref{ls:lm} trains a linear regression model. 
A random matrix \texttt{XY} is generated. 
The last column of \texttt{XY} contains the target variable, \texttt{Y}, and the remaining columns represent the features, \texttt{X}.
The code solves the linear system $Ax = b$, where the solution represents the coefficients of the linear regression model.
Figure~\ref{fig:ids-lm} shows the different components of the IDA in the linear regression model as realized by \daphnesched{}.
We use a \emph{a randomly generated matrix} as an input data set for the linear regression application.
\begin{lstlisting}[caption=Linear Regression algorithm in DaphneDSL~\cite{damme2022cidr}., label={ls:lm}, language=Python]
# Linear regression model training on random data.
# Data generation (in double precision).
XY = rand($numRows, $numCols, 0.0, 1.0, 1, -1);
# Extraction of X and y.
X = XY[, seq(0, as.si64($numCols) - 2, 1)];
y = XY[, seq(as.si64($numCols) - 1, as.si64($numCols) - 1, 1)];
# Normalization, standardization.
Xmeans = mean(X, 1);
Xstddev = stddev(X, 1);
X = (X - Xmeans) / Xstddev;
X = cbind(X, fill(1.0, nrow(X), 1));
A = syrk(X);
lambda = fill(0.001, ncol(X), 1);
A = A + diagMatrix(lambda);
b = gemv(X, y);
beta = solve(A, b);
\end{lstlisting}


\textbf{Target systems.} 
We used a two-socket Intel E5-2640 v4 (Broadwell), each with 10 cores, and a two-socket Intel Xeon Gold 6258R (Cascade Lake), each with 28 cores. 
The Broadwell and the Cascade Lake processors have 64 GB and 1500 GB RAM, respectively.

\begin{figure}
    \centering
    \subfloat[Connected components on Broadwell: The largest performance gain is by \texttt{MFSC}, $13.2\%$ faster than the default \texttt{STATIC} scheduling.]{\includegraphics[width=0.80\textwidth, keepaspectratio]{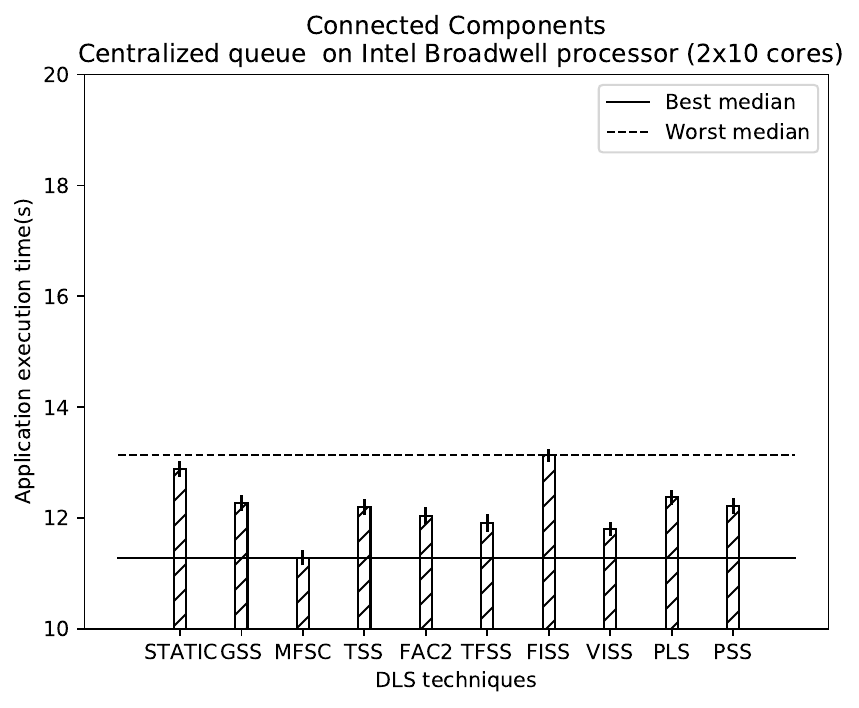}\label{fig:ib_centralized}}
   
    \centering
    \subfloat[Connected components on Cascade Lake: The largest performance gain is by \texttt{MFSC}, $8.3\%$ faster than the default \texttt{STATIC} scheduling.]
    {\includegraphics[width=0.80\textwidth, keepaspectratio]{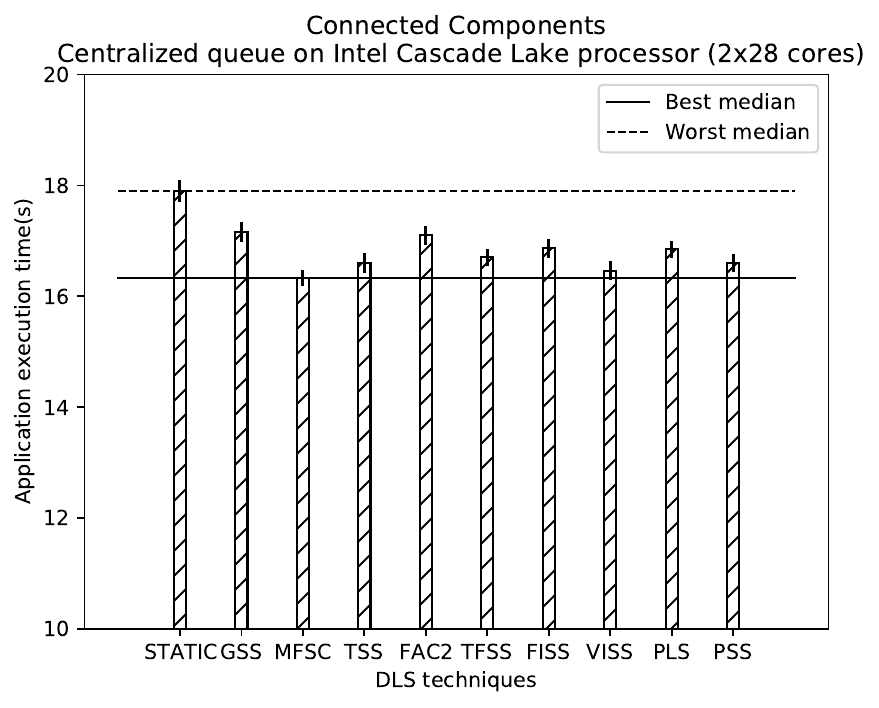}\label{fig:cl_centralized}}
    \caption{DAPHNE performance (execution time) of the connected components with one centralized work queue.}
\end{figure}

\textbf{Results.}
\AHE{In the experiments herein, we  compare various partitioning and assignment schemes against DAPHNE's default STATIC scheduling. In an earlier study, DAPHNE's STATIC scheduling outperformed established data analysis ecosystems~\cite{damme2022cidr}.} 
Figure~\ref{fig:ib_centralized} shows the performance of the connected components algorithm on the Broadwell processor. 
These results indicate that almost all scheduling techniques outperform the default \texttt{STATIC}, with \texttt{MFSC} bringing the largest performance gain of 13.2\%. 
The exception is \texttt{FISS}, with the application requiring 13.6s with \texttt{FISS} and 13.1s with \texttt{STATIC}. 
STATIC scheduling results in a few but coarse-grain tasks (one per worker). 
Such task partitioning is not optimal for the connected components algorithm with its highly sparse input data set (only 0.002\% are none-zeros)~\cite{leskovec2007dynamics}.  

The results in Figure~\ref{fig:cl_centralized} show a similar behavior, where \texttt{MFSC} delivers the largest performance gain of 8.3\% compared to the default \texttt{STATIC} scheduling. 
STATIC is still the least-performing scheduling technique on Cascade Lake.
The performance differences between all DLS techniques are minor (6\% difference in application execution times). 

The differences between the DLS techniques 
on Broadwell are almost double compared to the differences on Cascade Lake. 
This is because the Cascade Lake processor has more than double (56 cores) the number of cores on Broadwell (20 cores), which results in a more uniform work partitioning and assignment among the cores.
For instance, let us assume that the total number of tasks is 56. 
All DLS techniques on the Cascade Lake processor will perform similarly, indicating that regardless of the technique, there is a high possibility that each worker gets one task, and thus, resulting in a uniform task distribution. 
This work distribution does not guarantee load balancing, as if the variation in execution time between tasks is significant, the execution will be imbalanced regardless of the work distribution technique.


We also notice that, in general, the performance of the connected components algorithm on the Cascade Lake processor is lower than on the Broadwell processor. This is because of the performance cost of having a higher number of threads accessing locks simultaneously and the granularity of a single task execution time. 
To confirm this observation, we conducted certain experiments with \texttt{SS}, which considers the smallest task granularity of 1 task. We observed that the execution time explodes (tens of orders of magnitude) as many threads access the locks of the work queue, simultaneously. 
We omitted the results of \texttt{SS} on both platforms because adding these results in Figures~\ref{fig:ib_centralized} and~\ref{fig:cl_centralized} would occlude any insight.  

\begin{figure}
    \centering
    \subfloat[per core]{\includegraphics[clip,trim=0cm 3cm 0cm 1.7cm,width=\textwidth, keepaspectratio]{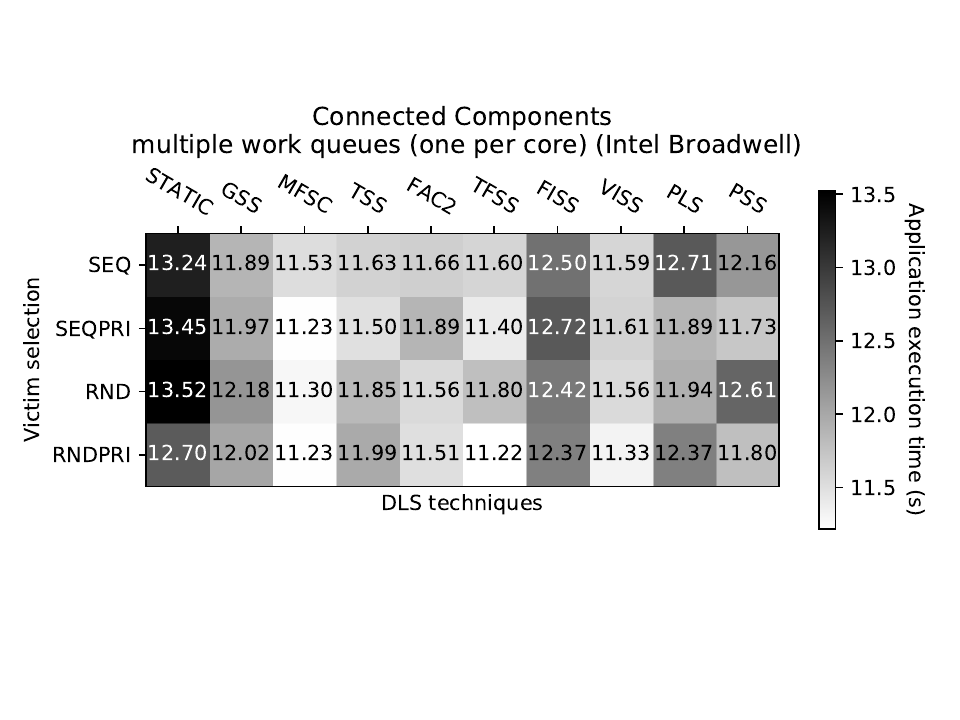}\label{fig:cc-bw-pcore}}
    
    \subfloat[per CPU]{\includegraphics[clip,trim=0cm 3cm 0cm 1.7Cm, width=\textwidth, keepaspectratio]{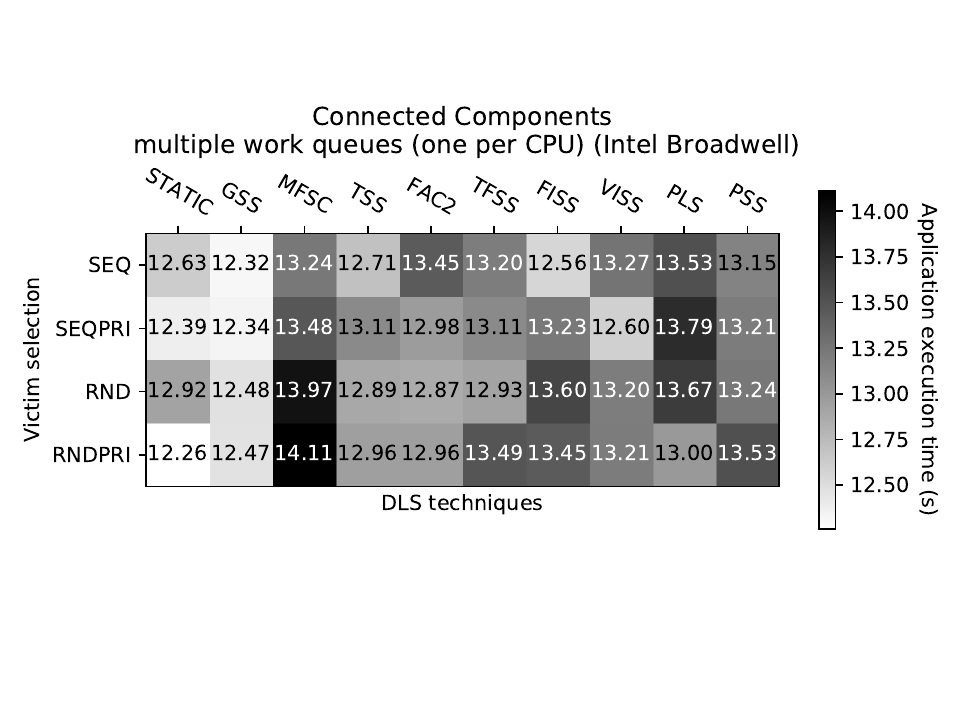}\label{fig:cc-bw-pcpu}} 
    \caption{Execution time of the connected components with DAPHNE, using multiple work queues - Intel Broadwell.}
    \label{fig:cc-broadwell-mq}
\end{figure}

Figure~\ref{fig:cc-broadwell-mq} shows the performance of the connected components algorithm when tasks are placed in \textit{multiple work queues}.
We observe that independent of the victim selection strategy, \texttt{STATIC} scheduling is not the lowest-performing technique, as in Figure~\ref{fig:cc-bw-pcpu}. 
In fact, for the \texttt{SEQPRI} strategy, \texttt{STATIC} is the highest-performing technique and even better than \texttt{STATIC} with a centralized queue (Fig.~\ref{fig:ib_centralized}).
This can be justified as follows: whenever the per CPU queue layout is chosen, \daphnesched{} initially partitions the data into a number of blocks that equals the total number of target CPUs. 
This step results in improved spatial data locality, meaning threads executing on cores within one CPU will have tasks from the same partition. 
In contrast, in Figure~\ref{fig:cc-bw-pcore}, we observe that \texttt{STATIC} scheduling is the lowest-performing technique, regardless of the victim selection strategy. We also observe that the performance of \texttt{STATIC} here is very close to the performance of \texttt{STATIC} in Figure~\ref{fig:ib_centralized} (one centralized work queue). 
This is because there is no pre-partitioning in both cases, i.e., workers (threads) arbitrarily obtain tasks in arbitrary order, and thus, threads executing on cores that share the lower level caches, or the NUMA domain may not have consecutive tasks (non-contiguous memory access). 

\begin{figure}
    \centering
    \subfloat[per core]{\includegraphics[clip,trim=0cm 3cm 0cm 1.7cm,width=\textwidth, keepaspectratio]{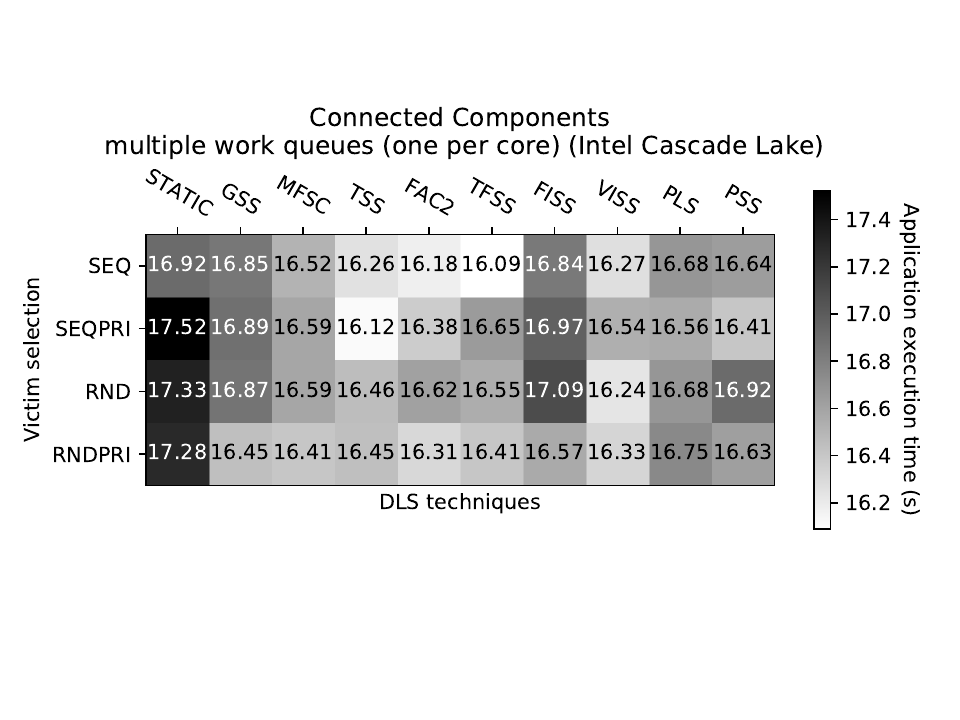}\label{fig:cc-cl-pcore}}
    
    \subfloat[per CPU]{\includegraphics[clip,trim=0cm 3cm 0cm 1.7cm,width=\textwidth, keepaspectratio]{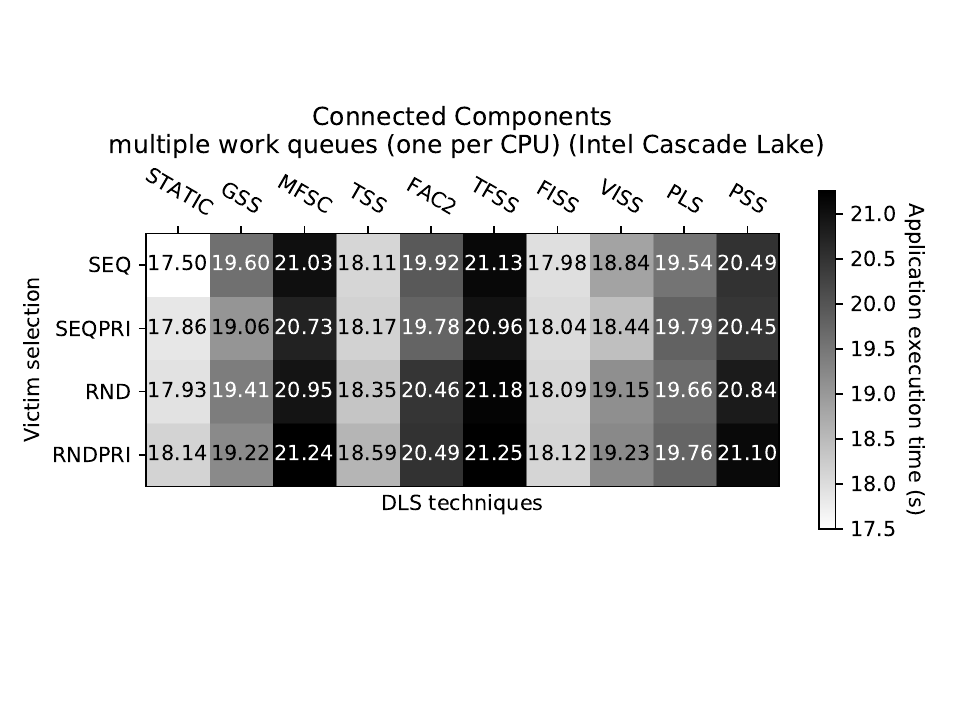}\label{fig:cc-cl-pcpu}}
    \caption{Execution time of the connected components with DAPHNE, using multiple work queues - Intel Cascade Lake.}
    \label{fig:cc-cl-mg}
\end{figure}

Figure~\ref{fig:cc-cl-pcpu} confirms this rationale, where \texttt{STATIC} is the highest-performing technique on the Cascade Lake processor independent of the victim selection strategy.  
   
\texttt{MFSC} shows an interesting behaviour in Figures~\ref{fig:cc-bw-pcore} and~\ref{fig:cc-bw-pcpu}.
Regardless of the victim selection strategy, in Figure~\ref{fig:cc-bw-pcore}, \texttt{MFSC} is among the highest-performing scheduling options, while in figure~\ref{fig:cc-bw-pcpu}, \texttt{MFSC} is among the lowest-performing techniques. 
This is due to \emph{lock contention}, i.e., in the case of \texttt{PERCPU}, the pre-partitioning results in decreasing the granularity of the tasks for \texttt{MFSC} by $\frac{1}{\#CPU}$. Thus, the execution time  of individual tasks becomes extremely short, and workers tend to simultaneously access the shared queue  more often resulting in high contention that hinders the potential of \texttt{MFSC}.

In Figure~\ref{fig:cc-bw-pcore}, the highest performance is observed with \texttt{TFSS} and the \texttt{RNDPRI} victim selection strategy.
In general, for the \texttt{PERCORE} work queue, \texttt{TFSS} scheduling, and \texttt{SEQ} victim selection yield the highest performance for the connected components on both processors.

\begin{figure}\label{fig:linreg-centralized}
    \centering
    \subfloat[Linear regression on Broadwell: The fastest scheduling technique is the default STATIC scheduling. The next fastest scheduling techniques are TSS and FISS, which are 16\% and 24\% worse than STATIC, respectively.]{\includegraphics[clip,trim= 0cm 0cm 0cm 0cm, width=0.8\textwidth, keepaspectratio]{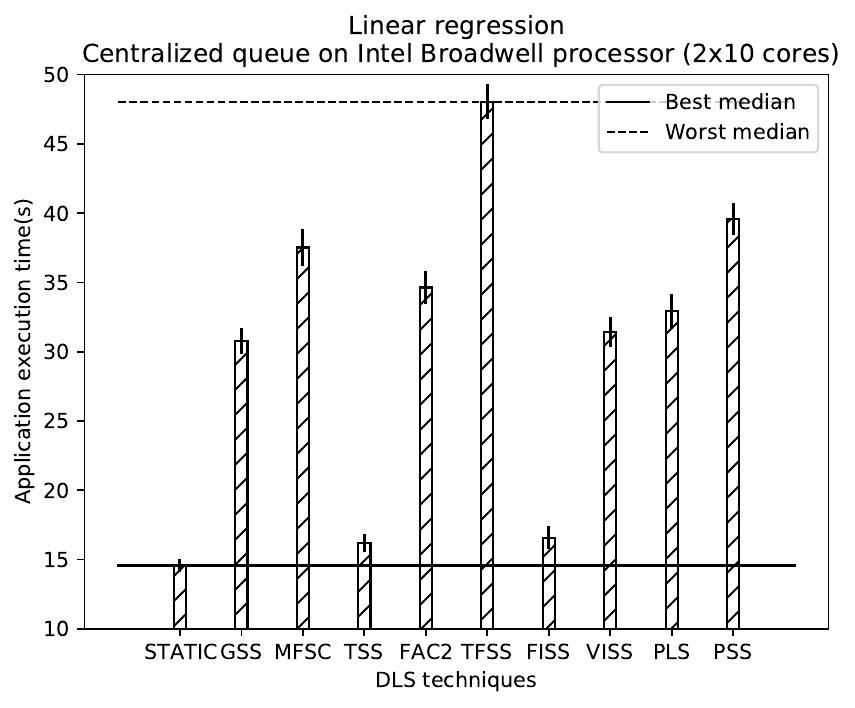}\label{fig:ib_lm_centralized}}
    \centering
    
    \subfloat[Linear regression on Cascade Lake: The fastest scheduling technique is the default STATIC scheduling. The next fastest scheduling techniques are TSS and FISS which are 50\% and 60\%  worse than STATIC, respectively.]{\includegraphics[clip,trim= 0cm 0cm 0cm 0cm, width=0.8\textwidth, keepaspectratio]{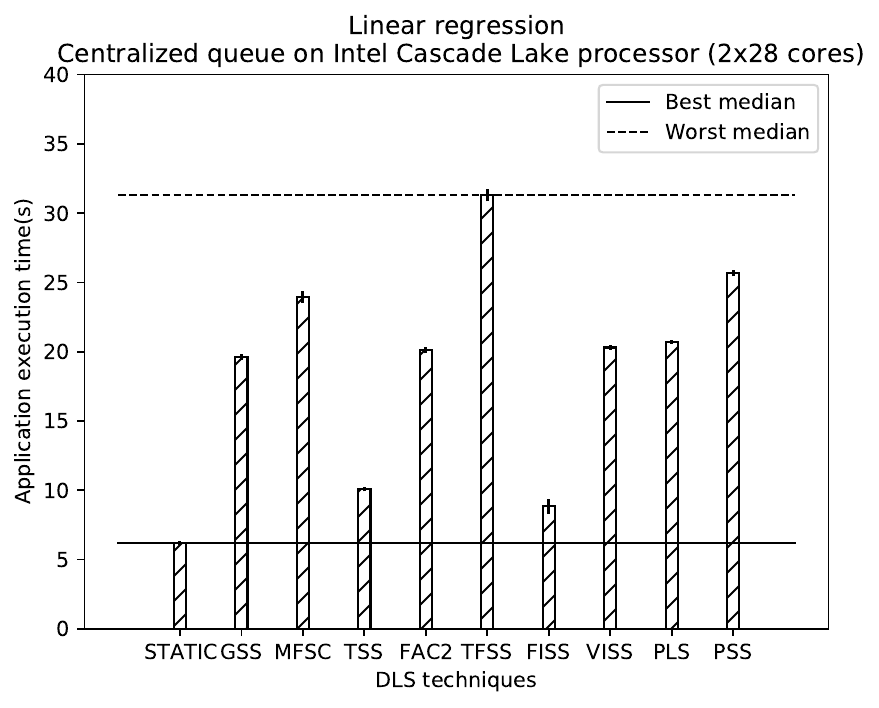}\label{fig:cl_lm_centralized}}
    \caption{Execution time of linear regression with DAPHNE, using a centralized work queue.}
\end{figure}

Figures~\ref{fig:ib_lm_centralized} and~\ref{fig:cl_lm_centralized} show
the performance of the linear regression model executed with DAPHNE on the two systems.
The highest performing scheduling technique in both cases is \texttt{STATIC}. 
All other techniques have much lower performance. 
For instance, in Figure~\ref{fig:ib_lm_centralized}, the linear regression executed in almost double the time with \texttt{MFSC}, \texttt{TFSS}, \texttt{PLS}, and \texttt{PSS} than \texttt{STATIC} scheduling.
This is because the input matrices to the linear regression application are \emph{dense}, which is the opposite of the connected components algorithm.
This leads to a rather load balanced execution. 
In such cases, employing DLS techniques may lead to performance degradation. 
Moreover, the scheduling overhead can artificially introduce load imbalance. 
This can happen, for instance, when a worker self-obtains a task and experiences contention on the work queue to retrieve it.
Since this contention may not occur throughout the entire application execution to every worker, the scheduling overhead may vary and contribute to the overall load imbalance.
In the case of linear regression, none of the DLS techniques achieve higher performance than \texttt{STATIC} due to the absence of load imbalance and associated scheduling overhead.
Figure~\ref{fig:cl_lm_centralized} shows similar results and performance observations on Cascade Lake to those of Broadwell.

\section{Conclusion and Future Work}
\label{sec:con}
DAPHNE is a software infrastructure that efficiently executes integrated data analysis pipelines, with \daphnesched{} at the core of its task scheduling. 
\daphnesched{} is designed to schedule IDA pipelines with both highly sparse (e.g., connected components algorithm) or highly dense (e.g., linear regression model) input data sets.  
\daphnesched{} provides a wide range of scheduling schemes for task partitioning and assignment, including self-scheduling and work-stealing. 
We showed that the number of workers and the work queues layout have a significant impact on the performance achievable with \daphnesched{}. Specifically, when multiple workers concurrently access a single work queue, the lock contention on the queue results in notable performance penalties, which can lead to performance degradation. 
For the two applications considered, the choice of the victim selection strategy is less critical than the choice of the work queue layout.

The lock contention on a single work queue seems to be a significant challenge that limits the potential of \daphnesched{} in certain cases. Further investigation of implementation alternatives is required to eliminate this limitation. 
We will consider atomic operations to access the queue instead of locks. 
Another limitation in the current version of \daphnesched{} is the lack of support for distributed-memory systems. 
There are ongoing efforts to support distributed scheduling with \daphnesched{} via MPI and RPC~\cite{d41}. 
Another important aspect is the multitude of scheduling options available in \daphnesched{} that renders the offline or online selection of the right scheduling option for an application-system pair very challenging. 
We plan to extend \daphnesched{} to support automatic selection of high performing scheduling algorithms and configurations.




 \section*{Acknowledgements}
 This project received funding from the European Union’s Horizon 2020 research and innovation programme under grant agreement No 957407 as DAPHNE.
 The authors also acknowledge Jonathan Giger and Gabrielle Poerwawinata for their earlier contributions to DaphneSched.

\bibliographystyle{IEEEtran}
\bibliography{daphnesched}

\end{document}